\documentclass[%
twocolumn,
superscriptaddress,
showpacs,
amsmath,amssymb,
aps,prl,nolongbibliography,
floatfix,
citeautoscript,
]{revtex4-2}

\usepackage[export]{adjustbox}
\usepackage[caption=false]{subfig}
\usepackage{graphicx}
\usepackage{xfrac}
\usepackage{bm}
\usepackage{xspace}
\usepackage{textcomp}

\setcitestyle{super}

\usepackage{color}

\definecolor{grey}{rgb}{.65,.65,.65}

\begin{document}

\newcommand{\ch}{$^{53}$Cr\xspace}
\newcommand{\co}{CrO$_2$\xspace}
\newcommand{\VV}{$V_{zz}$\xspace}
\newcommand{\cq}{$C_\mathrm{Q}$\xspace}
\newcommand{\tg}{$t_\mathrm{2g}$\xspace}
\newcommand{\xy}{$d_\mathrm{xy}$\xspace}
\newcommand{\xz}{$d_\mathrm{xz}$\xspace}
\newcommand{\yz}{$d_\mathrm{yz}$\xspace}

\newcommand{\texp}[1]{\ensuremath{\times 10^{#1}}}


\title{Electronic structure of \co probed by NMR and DFT}

\author{Vojt\v{e}ch Chlan}
\email{vojtech.chlan@mff.cuni.cz}
\affiliation{Charles University, Faculty of Mathematics and Physics, Department of Low Temperature Physics, V Holešovičkách 2, Prague 18000, Czech Republic}
\author{Anna A.\ Shmyreva}
\affiliation{Charles University, Faculty of Mathematics and Physics, Department of Low Temperature Physics, V Holešovičkách 2, Prague 18000, Czech Republic}\affiliation{St.\ Petersburg State University, Center for Magnetic Resonance, Universitetskiy pr.\ 26, St.\ Petersburg 198504, Russia}
\author{Helena \v{S}t\v{e}p\'{a}nkov\'{a}}
\affiliation{Charles University, Faculty of Mathematics and Physics, Department of Low Temperature Physics, V Holešovičkách 2, Prague 18000, Czech Republic}

\begin{abstract}

Electronic structure of ferromagnetic half-metal \co was studied by means of \ch nuclear magnetic resonance (NMR) spectroscopy and density functional theory (DFT). The measured NMR spectrum consists of three distinct spectral lines and is interpreted as a triplet arising due to electric quadrupole interaction. The observed NMR parameters agree well with those obtained from electronic structure calculations, corresponding to the presence of Cr$^{4+}$ with fully occupied localized $d_{xy}$ singlet and partially occupied degenerated $d_{xz}$ and $d_{yz}$ states, as required by ferromagnetic double exchange mechanism. With high accuracy the orbital occupations and valence states of all Cr atoms within the \co structure are found uniform. 
\end{abstract}

\maketitle
 
Chromium dioxide (\co) is a promising material for applications in spintronics owing to being a ferromagnetic half-metal \cite{deGroot1983, Coey2002}. The majority-spin electrons are metallic while there is a gap in the minority spin channel, leading to spin polarization close to 100~\% at the Fermi level \cite{Ji2001, Soulen1998, HuangExp2002}.

\co crystallizes in a tetragonal rutile structure\cite{Porta1972} (space group P$4_2$/$mnm$) where the two Cr atoms occupy octahedral 2$a$ sites at (0,0,0) and (\textonehalf,\textonehalf,\textonehalf). Chromium 3d orbitals are split into lower-lying \tg triplet, occupied by two electrons, and excited $e_\mathrm{g}$ doublet. The tetragonal distortion of the octahedron leads to a localized \xy singlet and degenerated \xz and \yz states hosting the remaining electron\cite{Sorantin1992,Lewis1997}. The microscopic mechanism of ferromagnetism in \co is double-exchange (DE), analogous to DE in manganites with a mixed valence state. However, in \co the localized and the itinerant \tg electrons at both Cr atoms are involved and in contrast to manganites this specific DE does not require the Cr atoms to be in different oxidation states \cite{Korotin1998,Schlottmann2003}. The metallic behavior of \co arises due to the dispersed chromium 3d states strongly hybridized with oxygen 2p states and stretching across the Fermi level \cite{Korotin1998}.

These basic features of the electronic structure are well reproduced by density functional theory (DFT) calculations already within the framework of the local-spin-density approximation (LSDA) or the generalized gradient approximation (GGA) \cite{Schwarz1986, Sorantin1992, Lewis1997, Kune2002}. Improved description of the orbital character, Cr orbital moments and other, more subtle properties require to include some electron correlation effects by the application of the LDA+U approach \cite{Korotin1998,Mazin1999,Laad2001,HuangDFT2002}, or by the dynamical mean field theory -- expectably a more adequate treatment of the half-metallic nature of \co \cite{Solovyev2015,Chioncel2007}.

The (half)metallic character of \co was confirmed directly by numerous experiments \cite{Ji2001,Soulen1998,HuangExp2002,Yates2007,Tsujioka1997}, however, the evidence of the specific orbital character of Cr, essential for the DE, is less straightforward to obtain experimentally. Optical measurements \cite{Singley1999,Huang2005,Stewart2009} as well as measurements of the x-ray absorption spectroscopy (XAS) and the x-ray magnetic circular dichroism (XMCD) \cite{Chang2005,HuangDFT2002,Stagarescu2000,Zimmermann2018} were interpreted in line with the picture of the band structure obtained by the calculations, however, other studies suggested different mechanism of DE, involving Cr atoms with mixed valence, by using XAS and XMCD \cite{Seong2018}, x-ray photoelectron spectroscopy \cite{HuangDFT2002}, or neutron powder diffraction \cite{Kodama2016}. Further such conclusion \cite{Shim2007,Takeda2012,Takeda2016} has been drawn from the nuclear magnetic resonance spectroscopy (NMR), which is particularly sensitive to the valence state and the local orbital arrangement of atoms in magnetic materials.

The first NMR measurements of \co by Nishihara et al.\cite{Nishihara1972} in 1972  were interpreted as originating from two different Cr species in the structure, because their \ch NMR spectra at 4.2~K clearly showed two distinct resonance lines at 26.3 and 36.7~MHz. Since then, it has been naturally assumed that there are two non-equivalent Cr species with markedly different hyperfine fields. Nishihara et al.\ assigned the line at 36.7~MHz to Cr in unperturbed \co structure while the second line was attributed to the presence of vacancies or similar structural imperfections. Later \ch NMR study \cite{Shim2007} assigned both \ch spectral lines as pertaining to the proper \co structure and the large difference in frequencies of the spectral lines was explained as a consequence of two different valence states Cr$^{(4\pm \delta)+}$; in analogy to DE mechanism in manganites, where two Mn species with different valence states can be observed in $^{55}$Mn NMR spectra \cite{Savosta1999,Savosta2003}. However, such an interpretation requires the difference $\delta$ in valence of the two Cr atoms to be relatively high ($\delta \sim 0.4$), which was not confirmed by other experimental methods or recognized in the calculations. Therefore, the spectrum was interpreted differently in later \ch NMR studies \cite{Takeda2012,Takeda2016} assuming that the pair of Cr atoms has much smaller difference in the valence states ($\delta \sim 0.03$), but rather differs in orbital occupations, allowing to explain the observed NMR spectrum by the anisotropic contributions to the hyperfine field at \ch nuclei. In very recent \ch NMR study including detailed analysis of the NMR relaxation times \cite{Piskunov2022} the authors arrived to a similar conclusion that the two observed \ch spectral lines correspond to Cr crystal sites with different local magnetic fields yet at the same time the Cr atoms may possess the same valence.

In this work we clear out these seeming contradictions by introducing new \ch NMR experiments where we detected (besides the two previously observed lines at 37.16 and 26.40~MHz) an additional spectral line at lower frequency. This line has been omitted in all previous NMR works on \co, and thus -- understandably -- the existing interpretations of the \ch NMR incorrectly assumed presence of two different Cr atoms in \co. In the view of the appearance of the third line, we reinterpret the \ch NMR spectrum in \co as a triplet owing to nuclear electric quadrupole interaction, which is rather strong here. Our NMR experiments thus show unambiguously that all Cr atoms in the \co structure are crystallographically and magnetically equivalent. Moreover, by using density functional theory (DFT) modeling we show that the observed NMR spectra correspond to the particular orbital arrangement, which is responsible for the specific DE mechanism in \co involving the localized and the itinerant \tg electrons at both Cr atoms in the unit cell\cite{Korotin1998,Schlottmann2003}.

\begin{figure}
	\centering
	\includegraphics[width=0.98\columnwidth]{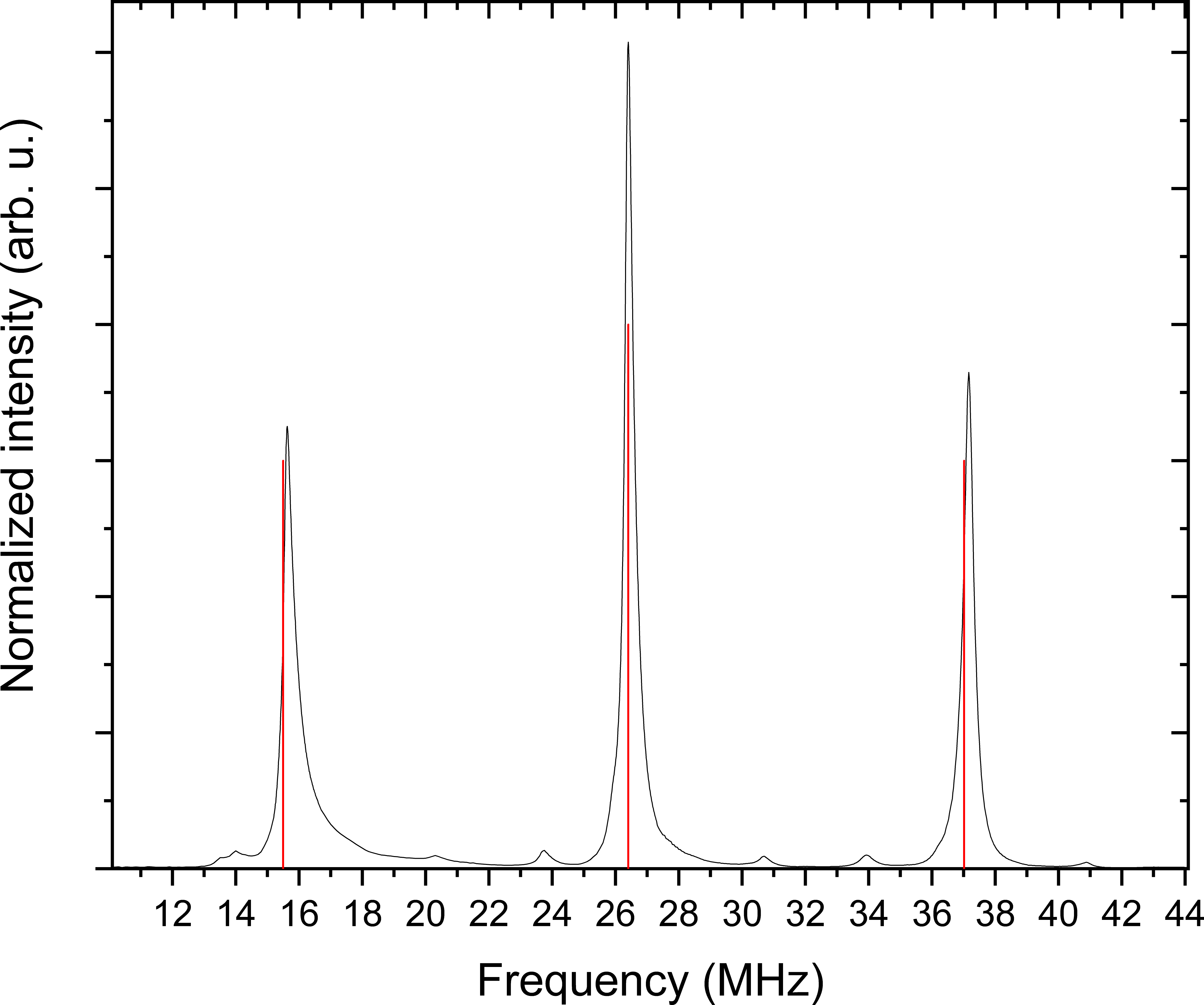}
	\caption{  \ch NMR spectrum at 4.2~K in zero external magnetic field. Red lines denote positions of the NMR lines based on the DFT-calculated \VV and $\eta$. Experimental spectrum was normalized by $f^2$ to reflect the linear frequency dependences of both the NMR probe inductance and the population differences of nuclear energy levels.} 
	\label{f1:spectrum}
\end{figure}

Studied \co powder sample was supplied by Sigma Aldrich (Magtrieve$^\mathrm{TM}$) and checked by x-ray diffraction as \co rutile structure with negligible traces of Cr$_2$O$_3$. Frequency swept \ch NMR spectrum was acquired in zero external magnetic field at temperature of 4.2~K. At each frequency step the NMR probe was properly tuned and matched, and Carr-Purcell-Meiboom-Gill pulse train was applied. All spin echos in the train were recorded and their sum Fourier transformed. The measured spectrum (Fig.~\ref{f1:spectrum}) consists of three intense spectral lines at 37.16, 26.40, and 15.63~MHz and several two orders of magnitude weaker lines spread over the displayed spectral region. We assign the intense triplet to the bulk \co phase, origin of the weaker signals is unknown, but we assume they arise from regions close to surface of the \co particles with less defined stoichiometry and crystalline arrangement. In this work we shall focus on the intense triplet only.

\begin{figure}
	\centering
	\includegraphics[width=0.98\columnwidth]{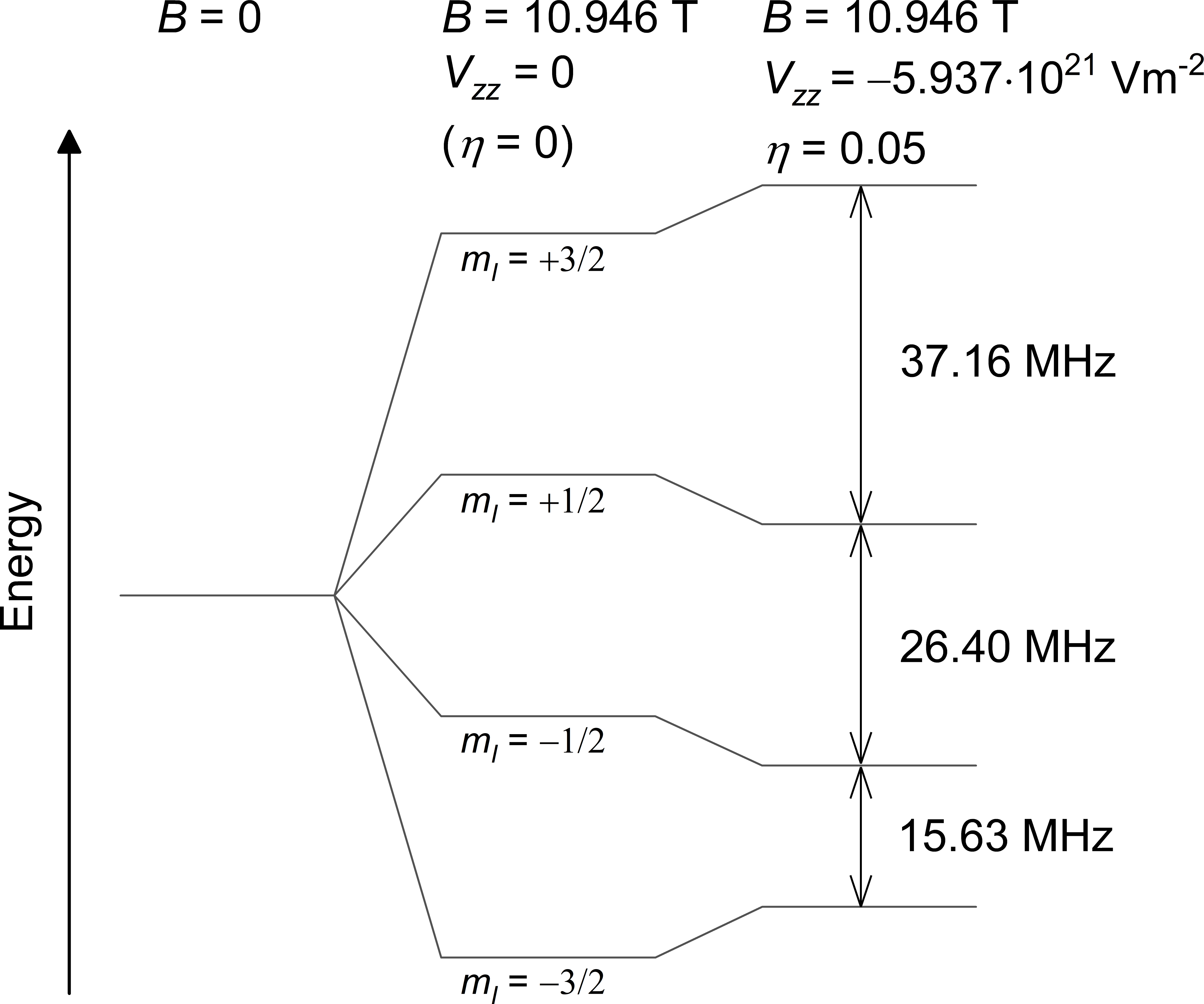}
	\caption{ Schematics of the effect of magnetic field $B$ and electric field gradient (EFG, parameters \VV and $\eta$) on the energy levels of the \ch nucleus (spin $I=\sfrac{3}{2}$). From left to right, the original degenerate ground state of the nucleus undergoes Zeeman splitting under the magnetic field $B$ into four levels (labeled by the magnetic quantum number $m_I$), and these are further shifted by the nuclear electric quadrupole interaction. Three transitions labeled with frequencies correspond to the three spectral lines observed in the NMR spectrum. Numerical values of $B$, $\left|V_{zz}\right|$, and $\eta$ were obtained from the fit of the experimental NMR spectrum. The sign of \VV was obtained from DFT calculations.} 
	\label{f2:diagram}
\end{figure}

Two lines at higher frequencies, 37.16 and 26.40~MHz, correspond to the lines documented in the previous NMR works \cite{Nishihara1972,Shim2007,Takeda2012,Takeda2016,Piskunov2022}, whereas the third line at 15.63~MHz is the one that has not been observed before. Nuclear spin number of \ch isotope $I=\sfrac{3}{2}$, and thus in a magnetic field $B$ the energy level of the nuclear ground state is Zeeman-split onto four equidistant stationary energy levels -- a situation that would lead to a single line in the NMR spectrum (gyromagnetic ratio\cite{Stone2005} of \ch nuclei $\frac{\gamma}{2\pi}=-2.4115\,\,\mathrm{MHz}\,\mathrm{T}^{-1}$). When additionally to the magnetic field an electric field gradient (EFG) is present at nuclei (nuclear quadrupole moment\cite{Ertmer1982} of \ch nuclei $Q=-150(50)\,\,\mathrm{milibarn}$, 1~milibarn = 10$^{-31}$~m$^{2}$), the four energy levels are further (unevenly) shifted due to the electric quadrupole interaction, which yields three different transitions observable in the NMR spectrum (see energy level diagram in Fig.~\ref{f2:diagram}). It is customary to label the energy levels by the magnetic quantum number $m_I$ of the original Zeeman eigenstates, and so in our case the line at 26.40 MHz arises due to the transition between levels $\left|-\sfrac{1}{2} \right>$ and $\left|\sfrac{1}{2} \right>$ (central transition, CT), while the lines at 37.16 and 15.63~MHz (satellite transitions, ST) correspond to the transitions $\left|\sfrac{1}{2} \right> \leftrightarrow \left|\sfrac{3}{2} \right>$ and $\left|-\sfrac{3}{2} \right> \leftrightarrow \left|-\sfrac{1}{2} \right>$, respectively. In case of \ch nucleus in \co without application of external magnetic field, the magnetic field $B$ at Cr nuclei results from hyperfine magnetic interaction of the \ch nuclear spin with the orbital and spin moments of electrons, mostly with the onsite Cr 3d states: directly (dipolar interaction) as well as mediated by spins of the s-states via Fermi contact interaction. Nonzero EFG appears at Cr sites owing to the local symmetry being lower than cubic (point group of the site symmetry is $mmm$).

\begin{figure}
	\centering
	\includegraphics[width=0.95\columnwidth]{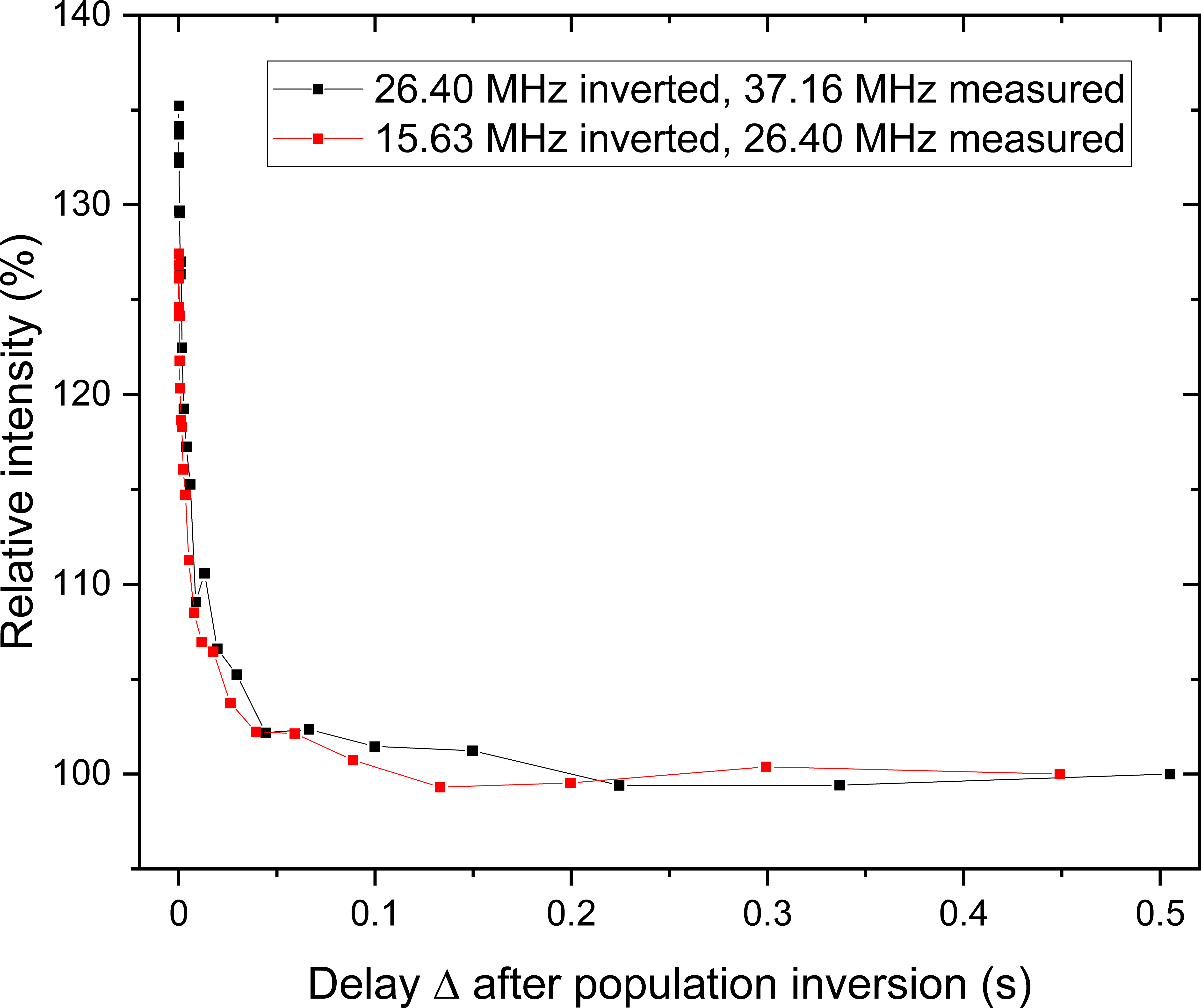}
	\caption{ Enhancement of the NMR signal in the double-resonance experiment. Population inversion at one transition causes an increase of the measured intensity at another transition of the \ch nuclear multiplet. 100~\% corresponds to intensity without an inverting pulse.} 
	\label{f3:double}
\end{figure}

We claim that the three observed spectral lines at 37.16, 26.40, and 15.63~MHz arise due to splitting of the spectrum of \ch in equivalent sites because of the electric quadrupole interaction. In such a case the three transitions are realized within one energy-level multiplet (Fig.~\ref{f2:diagram}), which can be unambiguously demonstrated by a double-resonance NMR experiment \cite{HAASE1998}. Energy level   $\left|\sfrac{1}{2}\right>$ is involved in two transitions: $\left|\sfrac{1}{2} \right> \leftrightarrow \left|\sfrac{3}{2}\right> $ producing one of the satellite lines (37.16~MHz), and $\left|-\sfrac{1}{2} \right> \leftrightarrow \left|\sfrac{1}{2}\right>$, corresponding to the CT at 26.40~MHz. Inducing population transfer between the levels of one transition changes the population difference for the other transition, which then affects the intensity of the corresponding spectral line. E.g., applying hard 180$^\circ$ radiofrequency (rf) pulse at frequency of 26.40~MHz (CT) inverts the populations of levels  $\left|-\sfrac{1}{2} \right> $ and $ \left|\sfrac{1}{2}\right>$ and the intensity of subsequently measured line at 37.16~MHz (ST) is enhanced in dependence on the delay $\Delta$ between the inverting 180$^\circ$ pulse at CT and the measuring echo-pulse sequence at ST (Fig.~\ref{f3:double}). Irradiation by the 180$^\circ$ rf pulse at 15.63~MHz and measuring at 26.40~MHz yields analogous result. The induced enhancement decreases with increasing delay $\Delta$ due to the nuclear spin-lattice relaxation process: in \co at 4.2~K the nuclear relaxation is relatively fast and $\Delta \sim 0.3$~s is sufficient for the inverted populations to revert back to the thermal equilibrium. The enhancement due to inversion is given by the Boltzmann distribution for the \ch multiplet displayed in Fig.~\ref{f2:diagram} and at temperature $T=4.2$~K the maximum enhancement equals ${\sim}171$ and ${\sim}158$~\% when applying inverting rf pulse at 26.40 and 15.63~MHz, respectively. This theoretical limit is not fully achieved in our experiments, though, most likely due to relatively large linewidth of the spectral lines compared to limited spectral bandwidth of the inverting pulse, and possibly also due to spin diffusion. Nonetheless, the observed enhancement directly proves that the three observed spectral lines are not three individual \ch species (with three different local magnetic fields), but belong to a triplet due to electric quadrupole interaction and thus originate from one type of Cr atoms with a single value of local magnetic field.

In order to confirm interpretation of the NMR spectrum by another independent method providing information on the electronic structure and hyperfine parameters, the electronic structure of \co was modelled within the DFT using the full-potential augmented plane-wave method implemented in WIEN2k \cite{Blaha2020}. In fact, our DFT calculations predicted the position of the line at 15.63~MHz prior performing the experiments, which emphasizes the importance of calculations in this field. Lattice parameters $a=4.4841$~\AA, $c=2.9745$~\AA, and oxygen parameter $u= 0.30168$ were fully relaxed within the space group P$4_2$/$mnm$. Perdew-Burke-Ernzerhof variant of the GGA exchange-correlation potential \cite{Perdew1996} was employed and the description of the electronic correlations was improved by the GGA+U approach applied to Cr 3d states with parameters $U_\mathrm{eff}=U-J=3.5$~eV and $J=0$~eV. Atomic sphere radii were chosen as 2.0 and 1.5~$a_0$ for Cr and O, respectively (Bohr unit $a_0 \sim 0.529$~\AA). We used computational parameters well converged with respect to EFG: a basis set of 1072 functions ($R_\mathrm{MT}K\mathrm{max} = 8.0$) and 2588 k-points (mesh $15\times 15\times 23$) in the irreducible part of the Brillouin zone. Spin-orbit interaction was introduced  for the semi-core and valence
electrons within the second variational method using the scalar-relativistic approximation \cite{MacDonald1980}.

The frequencies of the spectral lines of the \ch triplet are in general determined from eigenvalues of the spin Hamiltonian of electric quadrupole interaction and Zeemann interaction\cite{Abragam} for spin $I=\sfrac{3}{2}$:

\begin{eqnarray}\label{ham}
	H&=&\frac{eQV_{zz}}{4I(2I-1)}\left(3\hat{I}_z^2 - \hat{I}^2 + \frac{\eta}{2} (\hat{I}_+^2 + \hat{I}_-^2)   \right) + \\
	&+& \frac{\gamma B}{2} \left( \hat{I}_+e^{-i\varphi}\sin\vartheta + \hat{I}_-e^{i\varphi}\sin\vartheta  + 2\hat{I}_z \cos\vartheta\right) \nonumber
\end{eqnarray}
expressed using nuclear spin operators ($\hat{I}_z$, $\hat{I}$, $\hat{I}_\pm = \hat{I}_x\pm i\hat{I}_y$) within the principal axis system of the EFG tensor. The EFG tensor $V$ is defined by its largest principal component \VV, $\left| V_{zz} \right| \geq \left| V_{yy} \right| \geq \left| V_{xx} \right| $, and the asymmetry factor $\eta = \frac{V_{xx} - V_{yy}}{V_{zz}}$  ($0 \leq \eta \leq 1 $). $Q$ and $\gamma$ denote quadrupole moment and magnetogyric ratio of the nucleus in the ground state. Without an external magnetic field, the magnetic field $\bm{B}$ at the \ch nucleus is given by the hyperfine magnetic field $\bm{B}_\mathrm{hf}$. Orientation of $\bm{B}_\mathrm{hf}$ with respect to the main axes of EFG tensor is expressed via spherical angles $\vartheta$ and $\varphi$. For Cr and other 3d elements, the direction of $\bm{B}_\mathrm{hf}$ is antiparallel to the direction of atomic magnetic moment. In \co in zero external magnetic field both the direction of magnetization\cite{Yang2000} and the direction of \VV principal axis lie parallel to the tetragonal axis $c$, i.e., $\vartheta = 0$ and the dependence on $\varphi$ is removed from the Hamiltonian (Eq.~\ref{ham}). Parameters \VV and $\eta$ can be evaluated from the charge density calculated by DFT and are usually in a good agreement with experimental values for various compounds \cite{Cottenier2004,Zagorodniy2019,Choudhary2020}. Values of \VV and $\eta$ for \ch nuclei in \co were calculated as $V_{zz}= -5.95(5) \cdot 10^{21}$~$\mathrm{Vm}^{-2}$ and $\eta = 0.31(10)$ and arise predominately due to the Cr 3d states (d-d contribution), however, a weaker p-p contribution to \VV from the oxygen 2p states is also present. 
It is usual in NMR to express the strength of the electric quadrupole interaction by the quadrupole coupling constant, $C_Q=\frac{eQV_{zz}}{h} $, which here equals 21.6~MHz.

The hyperfine magnetic field $\bm{B}_\mathrm{hf}$ at Cr nuclei arises from the interaction of the nuclear spin with the orbital and spin moments of electrons surrounding the nucleus and is also obtainable from the calculations of electronic structure, however, for nuclei of transition-metal elements the Fermi contact term of the $\bm{B}_\mathrm{hf}$ is usually underestimated by the DFT calculations \cite{Novk2010}. In our case the calculated $B_\mathrm{hf} = 8.14(10)$~T at \ch nuclei is lower by about 26~\% than the experimentally observed value. 

Frequencies of the spectral lines obtained using the value of $B$ from experiment and the values of \VV and $\eta$ from the DFT calculation are compared to the experiment in Fig.~\ref{f1:spectrum}. The calculated \VV matches the experiment very well, the calculations underestimate the value of $\eta$, although the spectral shape is relatively insensitive to $\eta$ in this case. It should be noted, however, that the nuclear quadrupole moment $Q$ of \ch is known only relatively inaccurately \cite{Ertmer1982}, $Q = 150(50)$~milibarn. We presume that the accuracy of our DFT calculation of the EFG parameters is significantly higher than the accuracy of $Q$ and could be in principle used to refine its value.

The DFT calculations can also provide parameters describing the anisotropy of the hyperfine magnetic field, i.e., its dependence on the direction of magnetization, which can significantly influence the NMR spectrum in magnetic materials\cite{Chlan2011}. Our calculations show that for \co this anisotropy contribution is relatively large (it changes from $+2.2$~T for $\left[001\right]$ direction to $-5.0$~T for $\left[100\right]$ direction) and can be used to interpret NMR experiments on \co in external magnetic fields\cite{Nishihara1972,Shim2007,Takeda2012,Takeda2016}. Especially for the latter work\cite{Takeda2016}, where single crystal thin films were used, the explanation is relatively straightforward, since the frequency of the CT (26.40 MHz) is predominately influenced by the change of the local magnetic field due its anisotropy, whereas the ST at 37.16 is additionally strongly influenced by the change of the orientation of the local magnetic field (angles $\vartheta$ and $\varphi$ in Eq.~\ref{ham}) with respect to the EFG tensor.

We have shown that the observed \ch NMR spectrum consists of a triplet of lines due to electric quadrupole interaction, which corresponds to a single Cr species present in the \co structure. NMR parameters extracted from the experiment agree well with the calculated ones, indicating that the DFT calculations provide rather realistic description of \co electronic structure. In the following analysis we point out that this NMR-DFT correspondence is unique by showing that even a relatively low difference in the valence states or the orbital occupations of Cr atoms would lead to a notable change in the NMR spectrum. 

\begin{figure}
	\centering
	\includegraphics[width=0.5\columnwidth]{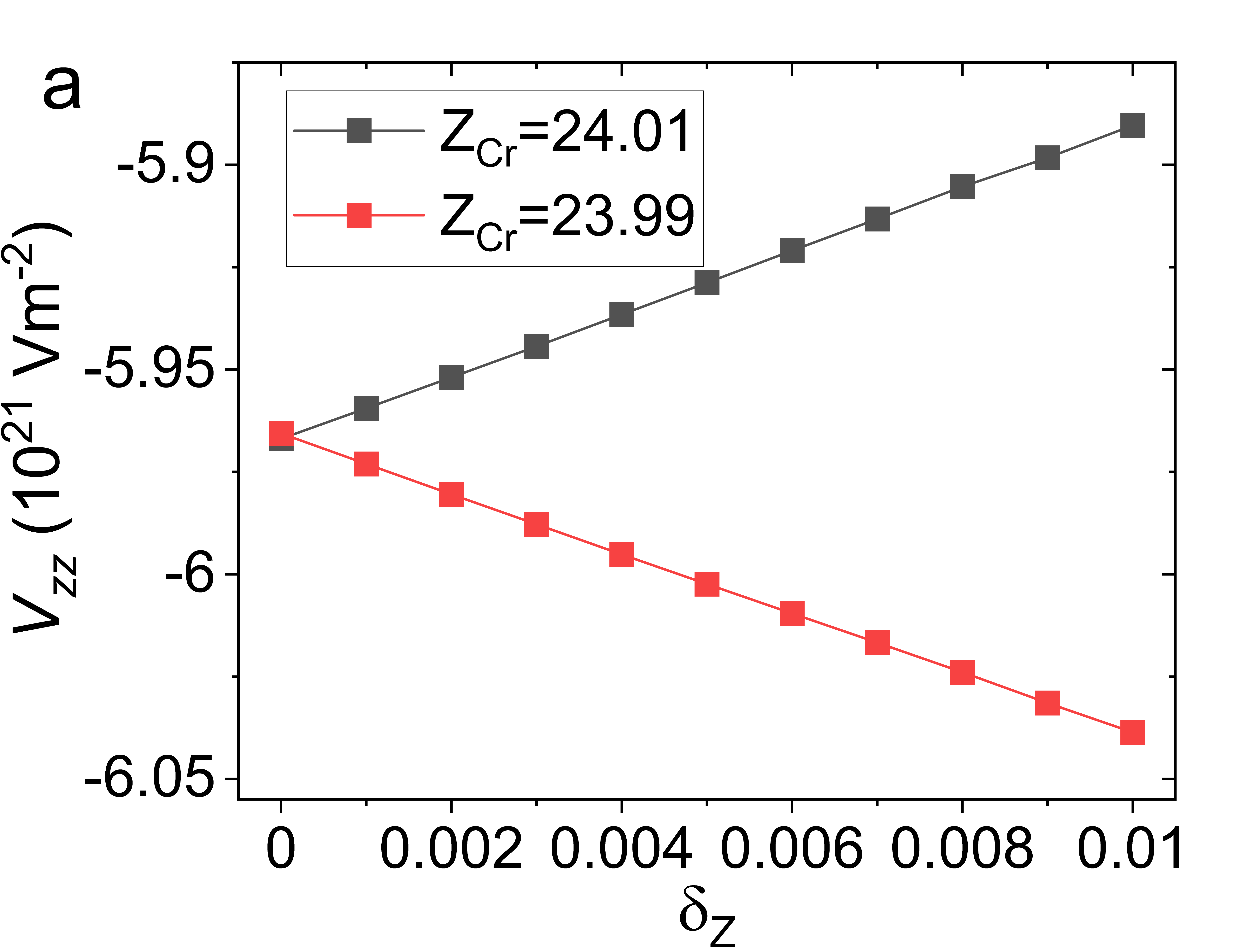}\includegraphics[width=0.5\columnwidth]{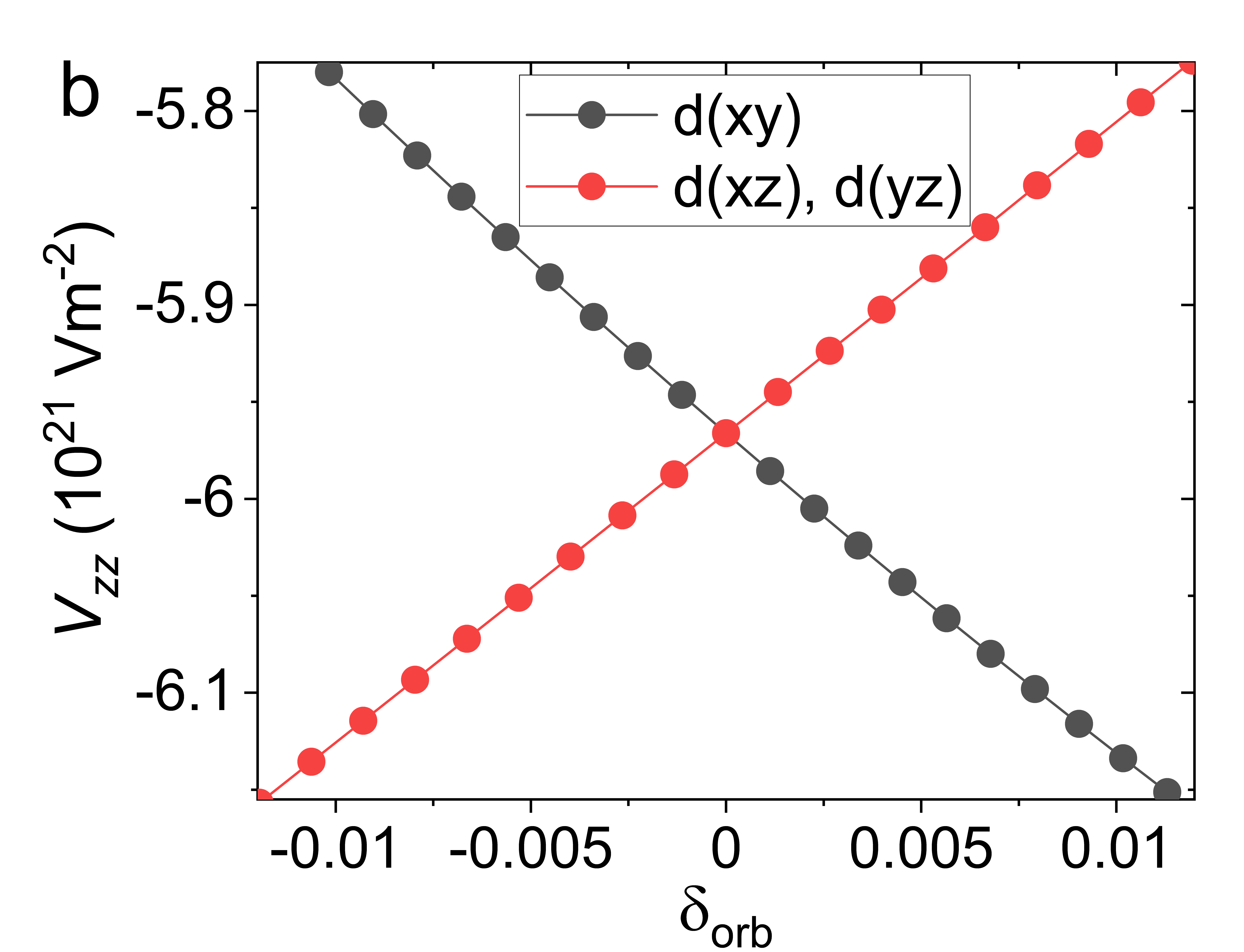}
	\caption{ The dependence of calculated \ch Vzz on the valence state of Cr atoms (a) and their orbital arrangement (b). } 
	\label{f4:modelling}
\end{figure}

In order to inspect the connection between \VV and the electronic state of Cr, we deliberately perturbed the calculated ground state and evaluated the dependence of \VV on the valence state of Cr atom and, independently, also on the orbital arrangement of Cr 3d states. The positive charges of the two Cr nuclei in the unit cell of \co are represented by the Coulomb potentials proportional to $Z=24$. Different valence states can be imposed on the two Cr atoms by modifying their atomic number $Z’ = 24 \pm \delta_Z$. Subsequent DFT calculation with such modified atomic numbers will reach a new ground state with the two Cr atoms possessing adequately differentiated valence states. The dependence of calculated \VV on the value of $\delta_Z$ is shown in Fig.~\ref{f4:modelling}a and we can estimate that $\delta_Z =0.01$, i.e., difference in the valence states of 0.02, would cause a difference of $0.15\cdot 10^{21}$~Vm$^{-2}$ between the values of \VV of the two Cr atoms, leading to a splitting of the ST lines about 2x larger than their linewidth, which would be well noticeable in the NMR experiment.

A simple illustration of how sensitively the EFG depends on the orbital distribution can be established by considering the d-d valence contribution to \VV and its proportionality to the "anisotropy count" of Cr 3d \cite{Blaha1988,AmbroschDraxl1989}:
\begin{equation}\label{aniz}
V_{zz} \simeq \Delta n_d=d_{xy}+d_{x^2-y^2}-\frac{1}{2}(d_{xz}+d_{yz})-d_{z^2}
\end{equation}

where $d_{xy}$, $d_{x^2-y^2}$, $d_{xz}$, $d_{yz}$, and $d_{z^2}$ are occupation numbers of the respective Cr d-states. \VV of Cr in \co is dominated by the d-d contribution and according to Eq.~\ref {aniz} the value of \VV should increase with increasing occupation of $d_{xy}$ or with decreasing occupation of $d_{xz}$ and $d_{yz}$. We may artificially perturb the occupations of Cr d-states by manually adjusting the corresponding occupation matrix in the calculations. Then, the applied orbital potential within the GGA+U framework pushes the occupations towards the desired state. From Fig.~\ref{f4:modelling}b we estimate that already a very small change, $\delta_\mathrm{orb}=0.01$, in the occupation of any of the \tg d-states would lead to an observable change of \VV by about $0.4\cdot10^{21}$~Vm$^{-2}$, producing a well visible frequency shift/splitting of the satellite lines in the NMR spectrum of $\sim700$~kHz. Given the observed widths of the satellite lines in the NMR spectrum we may conclude that the occupations of Cr d-states in \co are identical within the accuracy of 0.001.

In conclusion, the \ch NMR spectrum of \co measured at 4.2~K was interpreted on the basis of presence of strong nuclear electric quadrupole interaction. The calculations of electronic structure fully explain the observed NMR spectrum, which shows that the orbital occupations and valence states of both Cr sites in the unit cell of \co are identical and in line with the picture prevalent in the literature, i.e., the localized $d_{xy}$ singlet is occupied by one electron and the degenerated $d_{xz}$ and $d_{yz}$ states share the remaining electron of Cr$^{4+}$.

\begin{acknowledgments}
We thank R.\ Ku\v{z}el for the x-ray measurement. Computational resources were provided by the e-INFRA CZ project (ID:90254), supported by the Ministry of Education, Youth and Sports of the Czech Republic.
\end{acknowledgments}

\bibliography{cro2}

\end{document}